\documentclass[aps,prl,showpacs,notitlepage,amsmath,amssymb,floatfix,twocolumn,superscriptaddress]{revtex4-2}

\usepackage{amsmath} 
\usepackage{amssymb}
\usepackage{hyperref}
\usepackage[normalem]{ulem}
\usepackage{mathtools} 
\usepackage[dvipsnames]{xcolor}
\newcommand{\stkout}[1]{\ifmmode\text{\sout{\ensuremath{#1}}}\else\sout{#1}\fi}

\begin{document}
\title{Reconstructing Attractors with Autoencoders}
\author{F. Fainstein}
\affiliation{Universidad de Buenos Aires, Facultad de Ciencias Exactas y Naturales, Departamento de Física, Ciudad Universitaria, 1428 Buenos Aires, Argentina.}
\affiliation{CONICET - Universidad de Buenos Aires, Instituto de Física Interdisciplinaria y Aplicada (INFINA), Ciudad Universitaria, 1428 Buenos Aires, Argentina.}
\author{G. B. Mindlin}
\affiliation{Universidad de Buenos Aires, Facultad de Ciencias Exactas y Naturales, Departamento de Física, Ciudad Universitaria, 1428 Buenos Aires, Argentina.}
\affiliation{CONICET - Universidad de Buenos Aires, Instituto de Física Interdisciplinaria y Aplicada (INFINA), Ciudad Universitaria, 1428 Buenos Aires, Argentina.}
\author{P. Groisman}
\email[Corresponding author: ]{pgroisma@dm.uba.ar}
\affiliation{IMAS-CONICET and Departamento de Matemática, Facultad de Ciencias Exactas y Naturales, Universidad de Buenos Aires, Buenos Aires, Argentina.}

\begin{abstract} 
We show how to train an autoencoder to reconstruct an attractor from recorded footage, preserving the topology of the underlying phase space. This is explicitly demonstrated for the classic finite-amplitude Lorenz atmospheric convection problem.
\end{abstract}

\maketitle 

{\it Introduction.}--- In the natural sciences, many problems can be modeled in terms of dynamic systems. For those problems, the state of the system at each time corresponds to a point in a finite dimensional space, usually called the phase space. The temporal evolution is represented by a trajectory contained on a manifold $\mathcal{M}$. Elucidating the dynamics of the problem typically involves characterizing the attractor (which represents the long-time behavior) from the available data. This characterization can be done computing topological invariants, which act as fingerprints of the system \cite{guckenheimer2013nonlinear, gabo_libro_nonlinear}. However, a common challenge arises on how to reconstruct the attractor when we have access to a subset of the variables or, as in a video recording, when we do have access to a large set of measurements but the relevant variables are given by an unknown nonlinear transformation.

Currently, numerous approaches leverage the breakthrough of artificial intelligence to propose algorithms aimed at overcoming this challenge (see, for instance, Refs. \cite{ott_1, brunton_kutz_libro, champion_kutz_2019, agostini2020AEs, liu2021machine,floryan2022data, young2023deep}). The question arises if the reconstructed attractors preserve the topology of the original phase space of the problem. In this Letter we address this issue using an autoencoder, a prominent architecture in the realm of artificial neural networks \cite{hinton2006Science}. We focus on the case in which the data has been measured by an array of sensors that provide a temporally ordered sequence of frames capturing the spatiotemporal evolution of the system, and we discuss simple extensions for other cases. Through a formal mathematical argument, we propose a training loss function that guarantees that the reconstructed flow preserves the topology of the underlying manifold. This is explicitly demonstrated in the classic Lorenz model for finite-amplitude atmospheric convection \cite{lorenz1963deterministic}. 

\textit{The autoencoder.}--- An autoencoder is a type of artificial neural network specifically designed for dimensional reduction. It comprises an encoder ($E_{w}$) and a decoder ($D_{w}$). Both functions are implemented through layers of units (``neurons''), each one providing an output which is a nonlinear function of a linear combination of its inputs. The inputs are weighted sums of the outputs from the previous layer. The encoder maps the input data into a lower-dimensional space, and correspondingly, the decoder reconstructs the input from this encoded representation. The mid layer of the autoencoder, often referred to as the latent space, serves as a compressed representation of the data, where the most critical information about the input is condensed. Autoencoders can be trained with segments of time traces, sets of arbitrarily chosen features, or, as in our case, snapshots of our spatiotemporal data.

The training of a typical autoencoder proceeds as follows. Each frame is a set of $d_{\mathbf i}$ numbers (the number of pixels of each frame) and the weights in the network are chosen in a way that the sum of the differences between outputs and inputs, added over the whole dataset, is as small as possible. The interesting aspect of the method is that the number of units in the middle layer $d_{\mathbf l} < d_{\mathbf i}$ is chosen as small as possible. If the compression is pushed too much, there will be loss of information and the decoder will not be able to assign a unique output to each input frame. In other words, the successful recovery of a unique output for each input requires a unique representation of each element of the input in the latent space.

Carrying out this type of training, it was observed both by processing synthetic movies and others from actual experiments, that the encoder induces trajectories in the latent space whose topological organization is equivalent to the ones of the original phase space of the problem being filmed \cite{UribarriMindlin, fainstein2023reconstruction}. The reasons (if any) that guarantee this property to persist in general situations remain veiled, and it is the purpose of this Letter to propose a framework to solve this problem. 

\textit{Formal problem statement and proposal.}--- Let us assume that the  dynamics of our phenomenon is described by a dynamical system
\begin{equation}
\label{ydot}
\dot{y}(t) = \phi\left(y(t)\right), \qquad y\in \mathcal{M}.
\end{equation}
We do not have direct access to $y$ but our data is obtained through the application of a function $\alpha \in C^2(\mathcal{M}\times \mathcal{N}, \mathbb{R}^{d_\mathbf i})$ such that for every fixed $p$ in the manifold $\mathcal{N}$, $\alpha_p = \alpha(\cdot, p)$  is an embedding. The variable $p$ represents all the additional data that is not relevant for the problem but is saved when recording the movie. We will assume the existence of a function $\beta$ such that for every $p$ and $y$, $\beta(\alpha(y,p))=y$. This means that the video has been recorded in such a way that all the relevant variables can be recovered from the movie frame by frame, and $\beta$ is precisely the function that does so. For a given fixed $p$, we have $\beta^{-1} = \alpha_p$.  We do not have access neither to $\alpha$ nor $\beta$. We just observe $x(t)=\alpha_p(y(t))$ (but not $y(t)$) for a given background $p$ and trajectory $y(t)$.

The flow of a vector field $\phi$ on $\mathcal{M}$ consists on the transformations $g_{\phi}\colon\mathcal{M}\to\mathcal{M}$ converting every initial condition $y_{0}$ of Eq. \ref{ydot} at time $0$ on the value $g_{\phi}y_{0}$ of the solution at time $t$ \cite{arnold2012geometrical}. The objective of our work is to build a low dimensional representation of the flow from the movie (that we shall call $g_{\varphi}$), ensuring that it is topologically equivalent to the (unknown) flow ($g_{\phi}$). The proposal will be the representation obtained in the latent space of a specifically trained autoencoder.

Two flows $g_{\varphi} \colon \mathcal{M}_{1} \to \mathcal{M}_{1}$ and $g_{\Phi} \colon \mathcal{M}_{2} \to \mathcal{M}_{2}$ are topologically equivalent if there is a homeomorphism $h \colon \mathcal{M}_{1} \to \mathcal{M}_{2}$ mapping orbits of $g_{\varphi}$ to orbits of $g_{\Phi}$, preserving orientation of the orbits \cite{arnold2012geometrical}. Note that from our assumptions the movie and the original phenomenon are topologically equivalent so we can work directly with the recorded data. This means that if we want to ensure that $h=E_{w}$ provides a flow in the latent space topologically equivalent to $g_{\varphi}$, we need to show that: (1) the encoder $E_{w}$ is an homeomorphism, (2) the image of the trajectories in the input space are the solutions of a dynamical system (ruled by a vector field $\Phi$).

We consider that our data is dense enough, i.e. either our dynamics support a strange attractor so that we are sampling an orbit that is dense or we have samples of several trajectories that fill up $\mathcal{M}$. We will assume that we are in the first case and consider only one orbit, $x_{i}=x(t_{i})$, $t_{i}\in [0,T]$, with the labels ordered in time (i.e.: $t_{i+1} >t_i$ for every $i$). If we are in the second case we can proceed similarly but taking several trajectories instead of only one. We will guide the training with the loss function $\mathcal{L}= \lambda_{1} \mathcal{L}_{1} + \lambda_{2} \mathcal{L}_{2}$, composed by the usual mean squared error
\begin{equation}
\label{loss.1}
\mathcal{L}_1(w) = \sum_{x_i} \left | \left ( D_w \circ  E_w \right ) \left ( x_i \right ) - x_i\right |^2,
\end{equation}
\noindent and a second term given by
\begin{multline}
\label{loss.2}
\mathcal{L}_{2}(w) = \sum_{x_i} |(D_w \circ E_w ) ( x_{i+1} ) - ( D_w \circ E_w) ( x_{i} ) \\
 - ( x_{i+1} - x_i ) |^2.
\end{multline}
Here $w$ represents the weights of the network, the sum is over all the training set, and $\lambda_1$ and $\lambda_{2}$ are scaling factors. We will assume that the autoencoder generalizes, that means that $\mathcal{L}$ is small not just for the training set, but in all the input space $\alpha_{p}(\mathcal{M})\in \mathbb{R}^{d_\mathbf i}$ (for a discussion on this assumption see \cite{BelkinPNAS, BelkinPNAS2, BelkinActa}). This forces $E_{w}$ and $D_{w}$ to be continuous, and this means that $E_{w}$ is, asymptotically, an homeomorphism (with inverse $D_{w}$). 

What is not yet guaranteed is that the trajectories in the latent space $z(t)=E_{w}\left(x(t)\right)$ are in fact solutions of a dynamical system. The second term of our loss, $\mathcal{L}_{2}$, has been added precisely to enforce this. To see this, assuming our autoencoder is expresive enough, let us compute the vector field in the latent space by taking the directional derivative: 
\begin{equation}
\Phi\left( z \right) = \frac{\partial E_{w}}{\partial\varphi\left( x \right)}\left( x \right) \approx \frac{\partial E_{w}}{\partial\varphi\left( D_{w}\left( z \right) \right)}\left( D_{w}\left( z \right) \right).
\end{equation}
\noindent The second term states that
\begin{equation}
\frac{\partial\left( D_{w} \circ E_{w} \right)}{\partial\varphi\left( x \right)}\left( x \right) = \frac{\partial D_{w}}{\partial \Phi\left(z\right)} \left( z \right) \approx \varphi\left( x \right).
\end{equation}
\noindent This condition guarantees that nonzero velocities in the space of frames are not mapped into zero velocities in the latent space, that is $\Phi(z) \ne 0$ if $\varphi(x) \ne 0$. Note that it also penalizes strong spatial variations of $\Phi$ in the latent space, as one would expect for a smooth enough vector field defining a dynamical system. Therefore, in the limit as the size of the network goes to infinity, the phase flows $g_{\varphi}$ and $g_{\Phi}$ are topologically equivalent: the homeomorphism $E_{w}$ maps orbits of $\dot{x}=\varphi(x)$ to orbits of $\dot{z}=\Phi(z)$ homeomorphically and preserving orientation of the orbits.

\textit{Numerical work.}--- To numerically test our method, we generate a synthetic movie motivated by the classic model developed by Lorenz for an atmospheric convection problem \cite{lorenz1963deterministic}. Considering a layer of fluid of uniform depth $H$ and aspect ratio $a$, Lorenz proposes a modal decomposition for the stream function ($\psi$) and the departure of temperature from the non-convective state ($\theta$). He writes
\begin{equation}
\label{psi}
\psi = c_{1} X(t) \sin \left(\frac{\pi a x}{H} \right) \sin \left(\frac{\pi z}{H} \right),
\end{equation}
\begin{equation}
\label{theta}
\theta = c_{2} Y(t) \cos \left(\frac{\pi a x}{H}\right) \sin \left(\frac{\pi z}{H}\right) - c_{3} Z(t) \sin\left(\frac{2\pi z}{H}\right),
\end{equation}

\noindent where $X$, $Y$, $Z$ are functions of time alone, and $c_{1}$, $c_{2}$ and $c_{3}$ are constants. Here $x$ and $z$ denote the spatial coordinates of the problem. The dynamics of the modal amplitudes is determined by the famous dynamical system  
\begin{equation}
\begin{aligned}
\dot{X} &= \sigma\left( Y - X \right)\\
\dot{Y} &= {rX} - Y - {XZ}\\
\dot{Z} &= {XY} - {bZ}.
\end{aligned}
\end{equation}
By numerically integrating this system with $\sigma = 10$, $b = 8/3$ and $r = 28$, we generate temporal series with $40,000$ points (time step of $0.01$). We discard the first $1000$ points to avoid the transient state. Setting $a=H=1$, $c_{1}=c_{2}=2c_{3}=1/20$, and a spatial discretization of $0.025$, we use these time traces and Eqs. \ref{psi}-\ref{theta} to generate a sequence of $40\times40$ pixel frames describing the spatiotemporal pattern for $\psi$ and $\theta$. 

\begin{figure}[t] \centering
    \includegraphics[width=1\linewidth]{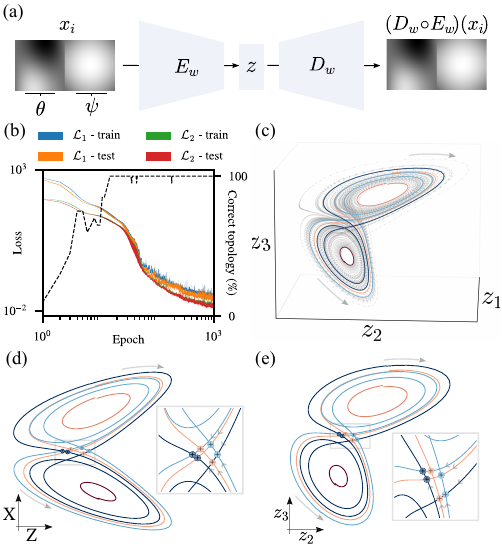} 
    \caption{(a) The autoencoder. Input data is encoded to and decoded from the latent space. (b) Evolution of both terms of our loss function, measured in the train and test datasets, left axis. Mean values over $20$ trained networks, with their standard error are shown. Evolution of the percentage of autoencoders with correct topology in the latent space as a function of the training epoch is shown in black dashed line, right axis. (c) Reconstructed flow in the latent space of an autoencoder. Five Lorenz unstable orbits and their linking numbers in phase space (d) and latent space (e). Dot color indicates which orbit is above at the crossing point.}
    \label{figura}
\end{figure}

To choose the proper training data set for our autoencoder, we note that not every movie is an embedding. For instance, the movie of a moving mass attached to a spring it is not. However, if we place a flag attached to the mass, it is, since the velocity could be recovered from each frame. In our case, we need information of the stream function and the temperature, so we train an autoencoder with a data set composed of inputs that contain both frames (see Fig. \ref{figura}(a)). The first 30000 frames are used for training, and the remaining 9000 frames for testing. 

The network architecture consists of initial and final fully connected layers of $40\times40\times2$ units. Intermediate fully connected layers containing $64$, $32$, and $16$ units encode the input to a three-dimensional latent space. Sine activation functions are used for all the layers except the latent and the output layers, that have no activation functions. The training process is conducted with a batch size of $600$ samples and $10^{3}$ epochs. The network is implemented in Python using PyTorch library (version 2.1.1). Adam optimization algorithm with learning rate $10^{-4}$, $\beta_{1}=0.9$ and $\beta_{2}=0.999$ is used, and scaling factors for the loss function are set to $\lambda_{1}=1$, $\lambda_{2}=50$ to ensure that both terms have similar magnitude. We do not perform any systematic search in the hyperparameter space to optimize the loss on the test set.

In Fig. \ref{figura}(b) we show the average evolution, over $20$ trained networks, of both terms of the cost function. In Fig. \ref{figura}(c) the reconstruction obtained in the latent space for one of the networks is shown. We remark that Fig \ref{figura}(c) is not the famous attractor given by Lorenz equations but the one reconstructed by the autoencoder.

To make a topological description, we use approximations of the unstable periodic orbits coexisting with the strange attractor. The way in which these orbits are intertwined is a fingerprint of the system. The organization of the orbits can be algebraically described by their relative rotation rates and linking numbers \cite{mindlin1990classification}. For two orbits of period one, one of period two, and two of period three, selected from the test set, we compute the relative linking numbers \cite{birman1983knotted}. In Figure \ref{figura} (d)-(e) we show the result in the original phase space and in the latent space, respectively. The evolution of the percentage of networks with correct topology during the training epochs is shown on the right axis of Figure \ref{figura}(b). All networks learnt the correct topology.

\textit{Conclusions.}--- In this study, we propose a method based on autoencoders to reconstruct the dynamics of a system from recorded footage. The method is applicable to any dataset which is an embedding, for instance, long-enough segments of time series data \cite{takens2006detecting}. Recent research suggested that such processing leads to a flow exhibiting a topological structure equivalent to the dynamical system governing the behavior
captured by the movie \cite{fainstein2023reconstruction}. Here, we showcase a modification to the loss function commonly used to guide the parameter tuning of an autoencoder, resulting in a trained network capable of reconstructing a flow topologically equivalent to the original one. It is not guaranteed to be an embedding: we are not tuning the encoder to have a differential that is invertible at any point but to be injective only in the direction of the flow. Yet, it turns out that this is enough to recover every topological property of the system.

Without this modification in the loss functions the autoencoder might still be able to produce a topologically equivalent dynamical system in the latent space, in a significant number of fitting runs. This might occur thanks to some kind of inducted bias/implicit regularization \cite{BelkinPNAS,BelkinPNAS2,BelkinActa} but it is not clear yet what kind of regularization might take place in particular runs. By adding the second term we force the autoencoder to have the regularization we need to ensure topological equivalence.

Recently, autoencoders were used to derive dynamical systems that could serve as models for data \cite{champion_kutz_2019}. The efforts provided algorithms for reconstructing vector fields in the latent space. Our work builds confidence on those efforts, by showing that properly trained, the flows in the latent space are topologically equivalent to the dynamics responsible for the data.

Over recent years, machine learning techniques have provided tools capable of solving extraordinarily complex problems, including the prediction of dynamics. Despite its apparent lack of necessity to illuminate the underlying mechanisms behind the learned dynamics, this work illustrates how the depth of the dynamics is embedded in some representation of the network. This study aligns with a series of observations made across various network architectures, and it would not be surprising if similar representation mechanisms could be formulated in those cases as well. This research contributes to our understanding of the nuanced interplay between artificial intelligence and dynamic systems, opening avenues for further exploration and application across diverse architectural structures.


\bibliography{template_PRL}

\begin{thebibliography}{20}%
\makeatletter
\providecommand \@ifxundefined [1]{%
 \@ifx{#1\undefined}
}%
\providecommand \@ifnum [1]{%
 \ifnum #1\expandafter \@firstoftwo
 \else \expandafter \@secondoftwo
 \fi
}%
\providecommand \@ifx [1]{%
 \ifx #1\expandafter \@firstoftwo
 \else \expandafter \@secondoftwo
 \fi
}%
\providecommand \natexlab [1]{#1}%
\providecommand \enquote  [1]{``#1''}%
\providecommand \bibnamefont  [1]{#1}%
\providecommand \bibfnamefont [1]{#1}%
\providecommand \citenamefont [1]{#1}%
\providecommand \href@noop [0]{\@secondoftwo}%
\providecommand \href [0]{\begingroup \@sanitize@url \@href}%
\providecommand \@href[1]{\@@startlink{#1}\@@href}%
\providecommand \@@href[1]{\endgroup#1\@@endlink}%
\providecommand \@sanitize@url [0]{\catcode `\\12\catcode `\$12\catcode
  `\&12\catcode `\#12\catcode `\^12\catcode `\_12\catcode `\%12\relax}%
\providecommand \@@startlink[1]{}%
\providecommand \@@endlink[0]{}%
\providecommand \url  [0]{\begingroup\@sanitize@url \@url }%
\providecommand \@url [1]{\endgroup\@href {#1}{\urlprefix }}%
\providecommand \urlprefix  [0]{URL }%
\providecommand \Eprint [0]{\href }%
\providecommand \doibase [0]{https://doi.org/}%
\providecommand \selectlanguage [0]{\@gobble}%
\providecommand \bibinfo  [0]{\@secondoftwo}%
\providecommand \bibfield  [0]{\@secondoftwo}%
\providecommand \translation [1]{[#1]}%
\providecommand \BibitemOpen [0]{}%
\providecommand \bibitemStop [0]{}%
\providecommand \bibitemNoStop [0]{.\EOS\space}%
\providecommand \EOS [0]{\spacefactor3000\relax}%
\providecommand \BibitemShut  [1]{\csname bibitem#1\endcsname}%
\let\auto@bib@innerbib\@empty
\bibitem [{\citenamefont {Guckenheimer}\ and\ \citenamefont
  {Holmes}(2013)}]{guckenheimer2013nonlinear}%
  \BibitemOpen
  \bibfield  {author} {\bibinfo {author} {\bibfnamefont {J.}~\bibnamefont
  {Guckenheimer}}\ and\ \bibinfo {author} {\bibfnamefont {P.}~\bibnamefont
  {Holmes}},\ }\href@noop {} {\emph {\bibinfo {title} {Nonlinear oscillations,
  dynamical systems, and bifurcations of vector fields}}},\ Vol.~\bibinfo
  {volume} {42}\ (\bibinfo  {publisher} {Springer Science \& Business Media},\
  \bibinfo {year} {2013})\BibitemShut {NoStop}%
\bibitem [{\citenamefont {Solari}\ \emph {et~al.}(1996)\citenamefont {Solari},
  \citenamefont {Natiello},\ and\ \citenamefont
  {Mindlin}}]{gabo_libro_nonlinear}%
  \BibitemOpen
  \bibfield  {author} {\bibinfo {author} {\bibfnamefont {H.~G.}\ \bibnamefont
  {Solari}}, \bibinfo {author} {\bibfnamefont {M.~A.}\ \bibnamefont
  {Natiello}},\ and\ \bibinfo {author} {\bibfnamefont {G.~B.}\ \bibnamefont
  {Mindlin}},\ }\href@noop {} {\emph {\bibinfo {title} {Nonlinear dynamics: a
  two-way trip from physics to math}}}\ (\bibinfo  {publisher} {Taylor \&
  Francis Group},\ \bibinfo {year} {1996})\BibitemShut {NoStop}%
\bibitem [{\citenamefont {Lu}\ \emph {et~al.}(2018)\citenamefont {Lu},
  \citenamefont {Hunt},\ and\ \citenamefont {Ott}}]{ott_1}%
  \BibitemOpen
  \bibfield  {author} {\bibinfo {author} {\bibfnamefont {Z.}~\bibnamefont
  {Lu}}, \bibinfo {author} {\bibfnamefont {B.~R.}\ \bibnamefont {Hunt}},\ and\
  \bibinfo {author} {\bibfnamefont {E.}~\bibnamefont {Ott}},\ }\bibfield
  {title} {\bibinfo {title} {Attractor reconstruction by machine learning},\
  }\href {https://doi.org/10.1063/1.5039508} {\bibfield  {journal} {\bibinfo
  {journal} {Chaos}\ }\textbf {\bibinfo {volume} {28}} (\bibinfo {year}
  {2018})}\BibitemShut {NoStop}%
\bibitem [{\citenamefont {Brunton}\ and\ \citenamefont
  {Kutz}(2019)}]{brunton_kutz_libro}%
  \BibitemOpen
  \bibfield  {author} {\bibinfo {author} {\bibfnamefont {S.~L.}\ \bibnamefont
  {Brunton}}\ and\ \bibinfo {author} {\bibfnamefont {J.~N.}\ \bibnamefont
  {Kutz}},\ }\href@noop {} {\emph {\bibinfo {title} {Data-Driven Science and
  Engineering: Machine Learning, Dynamical Systems, and Control}}}\ (\bibinfo
  {publisher} {Cambridge University Press, Cambridge, England},\ \bibinfo
  {year} {2019})\BibitemShut {NoStop}%
\bibitem [{\citenamefont {Champion}\ \emph {et~al.}(2019)\citenamefont
  {Champion}, \citenamefont {Lusch}, \citenamefont {Kutz},\ and\ \citenamefont
  {Brunton}}]{champion_kutz_2019}%
  \BibitemOpen
  \bibfield  {author} {\bibinfo {author} {\bibfnamefont {K.}~\bibnamefont
  {Champion}}, \bibinfo {author} {\bibfnamefont {B.}~\bibnamefont {Lusch}},
  \bibinfo {author} {\bibfnamefont {J.~N.}\ \bibnamefont {Kutz}},\ and\
  \bibinfo {author} {\bibfnamefont {S.~L.}\ \bibnamefont {Brunton}},\
  }\bibfield  {title} {\bibinfo {title} {Data-driven discovery of coordinates
  and governing equations},\ }\href
  {https://doi.org/https://doi.org/10.1073/pnas.1906995116} {\bibfield
  {journal} {\bibinfo  {journal} {Proc. Natl. Acad. Sci. USA}\ }\textbf
  {\bibinfo {volume} {116}},\ \bibinfo {pages} {22445} (\bibinfo {year}
  {2019})}\BibitemShut {NoStop}%
\bibitem [{\citenamefont {Agostini}(2020)}]{agostini2020AEs}%
  \BibitemOpen
  \bibfield  {author} {\bibinfo {author} {\bibfnamefont {L.}~\bibnamefont
  {Agostini}},\ }\bibfield  {title} {\bibinfo {title} {Exploration and
  prediction of fluid dynamical systems using auto-encoder technology},\
  }\bibfield  {journal} {\bibinfo  {journal} {Physics of Fluids}\ }\textbf
  {\bibinfo {volume} {32}},\ \href
  {https://doi.org/https://doi.org/10.1063/5.0012906}
  {https://doi.org/10.1063/5.0012906} (\bibinfo {year} {2020})\BibitemShut
  {NoStop}%
\bibitem [{\citenamefont {Liu}\ and\ \citenamefont
  {Tegmark}(2021)}]{liu2021machine}%
  \BibitemOpen
  \bibfield  {author} {\bibinfo {author} {\bibfnamefont {Z.}~\bibnamefont
  {Liu}}\ and\ \bibinfo {author} {\bibfnamefont {M.}~\bibnamefont {Tegmark}},\
  }\bibfield  {title} {\bibinfo {title} {Machine learning conservation laws
  from trajectories},\ }\href
  {https://doi.org/https://doi.org/10.1103/PhysRevLett.126.180604} {\bibfield
  {journal} {\bibinfo  {journal} {Physical Review Letters}\ }\textbf {\bibinfo
  {volume} {126}},\ \bibinfo {pages} {180604} (\bibinfo {year}
  {2021})}\BibitemShut {NoStop}%
\bibitem [{\citenamefont {Floryan}\ and\ \citenamefont
  {Graham}(2022)}]{floryan2022data}%
  \BibitemOpen
  \bibfield  {author} {\bibinfo {author} {\bibfnamefont {D.}~\bibnamefont
  {Floryan}}\ and\ \bibinfo {author} {\bibfnamefont {M.~D.}\ \bibnamefont
  {Graham}},\ }\bibfield  {title} {\bibinfo {title} {Data-driven discovery of
  intrinsic dynamics},\ }\href
  {https://doi.org/https://doi.org/10.1038/s42256-022-00575-4} {\bibfield
  {journal} {\bibinfo  {journal} {Nature Machine Intelligence}\ }\textbf
  {\bibinfo {volume} {4}},\ \bibinfo {pages} {1113} (\bibinfo {year}
  {2022})}\BibitemShut {NoStop}%
\bibitem [{\citenamefont {Young}\ and\ \citenamefont
  {Graham}(2023)}]{young2023deep}%
  \BibitemOpen
  \bibfield  {author} {\bibinfo {author} {\bibfnamefont {C.~D.}\ \bibnamefont
  {Young}}\ and\ \bibinfo {author} {\bibfnamefont {M.~D.}\ \bibnamefont
  {Graham}},\ }\bibfield  {title} {\bibinfo {title} {Deep learning delay
  coordinate dynamics for chaotic attractors from partial observable data},\
  }\href {https://doi.org/https://doi.org/10.1103/PhysRevE.107.034215}
  {\bibfield  {journal} {\bibinfo  {journal} {Physical Review E}\ }\textbf
  {\bibinfo {volume} {107}},\ \bibinfo {pages} {034215} (\bibinfo {year}
  {2023})}\BibitemShut {NoStop}%
\bibitem [{\citenamefont {Hinton}\ and\ \citenamefont
  {Salakhutdinov}(2006)}]{hinton2006Science}%
  \BibitemOpen
  \bibfield  {author} {\bibinfo {author} {\bibfnamefont {G.~E.}\ \bibnamefont
  {Hinton}}\ and\ \bibinfo {author} {\bibfnamefont {R.~R.}\ \bibnamefont
  {Salakhutdinov}},\ }\bibfield  {title} {\bibinfo {title} {Reducing the
  dimensionality of data with neural networks},\ }\href
  {https://doi.org/10.1126/science.1127647} {\bibfield  {journal} {\bibinfo
  {journal} {Science}\ }\textbf {\bibinfo {volume} {313}},\ \bibinfo {pages}
  {504} (\bibinfo {year} {2006})}\BibitemShut {NoStop}%
\bibitem [{\citenamefont {Lorenz}(1963)}]{lorenz1963deterministic}%
  \BibitemOpen
  \bibfield  {author} {\bibinfo {author} {\bibfnamefont {E.~N.}\ \bibnamefont
  {Lorenz}},\ }\bibfield  {title} {\bibinfo {title} {Deterministic nonperiodic
  flow},\ }\href@noop {} {\bibfield  {journal} {\bibinfo  {journal} {Journal of
  atmospheric sciences}\ }\textbf {\bibinfo {volume} {20}},\ \bibinfo {pages}
  {130} (\bibinfo {year} {1963})}\BibitemShut {NoStop}%
\bibitem [{\citenamefont {Uribarri}\ and\ \citenamefont
  {Mindlin}(2020)}]{UribarriMindlin}%
  \BibitemOpen
  \bibfield  {author} {\bibinfo {author} {\bibfnamefont {G.}~\bibnamefont
  {Uribarri}}\ and\ \bibinfo {author} {\bibfnamefont {G.~B.}\ \bibnamefont
  {Mindlin}},\ }\bibfield  {title} {\bibinfo {title} {The structure of
  reconstructed flows in latent spaces},\ }\href
  {https://doi.org/10.1063/5.0013714} {\bibfield  {journal} {\bibinfo
  {journal} {Chaos}\ }\textbf {\bibinfo {volume} {30}},\ \bibinfo {pages}
  {093109, 8} (\bibinfo {year} {2020})}\BibitemShut {NoStop}%
\bibitem [{\citenamefont {Fainstein}\ \emph {et~al.}(2023)\citenamefont
  {Fainstein}, \citenamefont {Catoni}, \citenamefont {Elemans},\ and\
  \citenamefont {Mindlin}}]{fainstein2023reconstruction}%
  \BibitemOpen
  \bibfield  {author} {\bibinfo {author} {\bibfnamefont {F.}~\bibnamefont
  {Fainstein}}, \bibinfo {author} {\bibfnamefont {J.}~\bibnamefont {Catoni}},
  \bibinfo {author} {\bibfnamefont {C.~P.}\ \bibnamefont {Elemans}},\ and\
  \bibinfo {author} {\bibfnamefont {G.~B.}\ \bibnamefont {Mindlin}},\
  }\bibfield  {title} {\bibinfo {title} {The reconstruction of flows from
  spatiotemporal data by autoencoders},\ }\href
  {https://doi.org/https://doi.org/10.1016/j.chaos.2023.114115} {\bibfield
  {journal} {\bibinfo  {journal} {Chaos, Solitons \& Fractals}\ }\textbf
  {\bibinfo {volume} {176}},\ \bibinfo {pages} {114115} (\bibinfo {year}
  {2023})}\BibitemShut {NoStop}%
\bibitem [{\citenamefont {Arnold}(2012)}]{arnold2012geometrical}%
  \BibitemOpen
  \bibfield  {author} {\bibinfo {author} {\bibfnamefont {V.~I.}\ \bibnamefont
  {Arnold}},\ }\href@noop {} {\emph {\bibinfo {title} {Geometrical methods in
  the theory of ordinary differential equations}}},\ Vol.\ \bibinfo {volume}
  {250}\ (\bibinfo  {publisher} {Springer Science \& Business Media},\ \bibinfo
  {year} {2012})\BibitemShut {NoStop}%
\bibitem [{\citenamefont {Belkin}\ \emph {et~al.}(2019)\citenamefont {Belkin},
  \citenamefont {Hsu}, \citenamefont {Ma},\ and\ \citenamefont
  {Mandal}}]{BelkinPNAS}%
  \BibitemOpen
  \bibfield  {author} {\bibinfo {author} {\bibfnamefont {M.}~\bibnamefont
  {Belkin}}, \bibinfo {author} {\bibfnamefont {D.}~\bibnamefont {Hsu}},
  \bibinfo {author} {\bibfnamefont {S.}~\bibnamefont {Ma}},\ and\ \bibinfo
  {author} {\bibfnamefont {S.}~\bibnamefont {Mandal}},\ }\bibfield  {title}
  {\bibinfo {title} {Reconciling modern machine-learning practice and the
  classical bias-variance trade-off},\ }\href
  {https://doi.org/10.1073/pnas.1903070116} {\bibfield  {journal} {\bibinfo
  {journal} {Proc. Natl. Acad. Sci. USA}\ }\textbf {\bibinfo {volume} {116}},\
  \bibinfo {pages} {15849} (\bibinfo {year} {2019})}\BibitemShut {NoStop}%
\bibitem [{\citenamefont {Radhakrishnan}\ \emph {et~al.}(2020)\citenamefont
  {Radhakrishnan}, \citenamefont {Belkin},\ and\ \citenamefont
  {Uhler}}]{BelkinPNAS2}%
  \BibitemOpen
  \bibfield  {author} {\bibinfo {author} {\bibfnamefont {A.}~\bibnamefont
  {Radhakrishnan}}, \bibinfo {author} {\bibfnamefont {M.}~\bibnamefont
  {Belkin}},\ and\ \bibinfo {author} {\bibfnamefont {C.}~\bibnamefont
  {Uhler}},\ }\bibfield  {title} {\bibinfo {title} {Overparameterized neural
  networks implement associative memory},\ }\href
  {https://doi.org/10.1073/pnas.2005013117} {\bibfield  {journal} {\bibinfo
  {journal} {Proc. Natl. Acad. Sci. USA}\ }\textbf {\bibinfo {volume} {117}},\
  \bibinfo {pages} {27163} (\bibinfo {year} {2020})}\BibitemShut {NoStop}%
\bibitem [{\citenamefont {Belkin}(2021)}]{BelkinActa}%
  \BibitemOpen
  \bibfield  {author} {\bibinfo {author} {\bibfnamefont {M.}~\bibnamefont
  {Belkin}},\ }\bibfield  {title} {\bibinfo {title} {Fit without fear:
  remarkable mathematical phenomena of deep learning through the prism of
  interpolation},\ }\href {https://doi.org/10.1017/S0962492921000039}
  {\bibfield  {journal} {\bibinfo  {journal} {Acta Numerica}\ }\textbf
  {\bibinfo {volume} {30}},\ \bibinfo {pages} {203} (\bibinfo {year}
  {2021})}\BibitemShut {NoStop}%
\bibitem [{\citenamefont {Mindlin}\ \emph {et~al.}(1990)\citenamefont
  {Mindlin}, \citenamefont {Hou}, \citenamefont {Solari}, \citenamefont
  {Gilmore},\ and\ \citenamefont {Tufillaro}}]{mindlin1990classification}%
  \BibitemOpen
  \bibfield  {author} {\bibinfo {author} {\bibfnamefont {G.~B.}\ \bibnamefont
  {Mindlin}}, \bibinfo {author} {\bibfnamefont {X.-J.}\ \bibnamefont {Hou}},
  \bibinfo {author} {\bibfnamefont {H.~G.}\ \bibnamefont {Solari}}, \bibinfo
  {author} {\bibfnamefont {R.}~\bibnamefont {Gilmore}},\ and\ \bibinfo {author}
  {\bibfnamefont {N.}~\bibnamefont {Tufillaro}},\ }\bibfield  {title} {\bibinfo
  {title} {Classification of strange attractors by integers},\ }\href
  {https://doi.org/https://doi.org/10.1103/PhysRevLett.64.2350} {\bibfield
  {journal} {\bibinfo  {journal} {Physical Review Letters}\ }\textbf {\bibinfo
  {volume} {64}},\ \bibinfo {pages} {2350} (\bibinfo {year}
  {1990})}\BibitemShut {NoStop}%
\bibitem [{\citenamefont {Birman}\ and\ \citenamefont
  {Williams}(1983)}]{birman1983knotted}%
  \BibitemOpen
  \bibfield  {author} {\bibinfo {author} {\bibfnamefont {J.~S.}\ \bibnamefont
  {Birman}}\ and\ \bibinfo {author} {\bibfnamefont {R.~F.}\ \bibnamefont
  {Williams}},\ }\bibfield  {title} {\bibinfo {title} {Knotted periodic orbits
  in dynamical systems i: Lorenz’s equations},\ }\href
  {https://doi.org/https://doi.org/10.1016/0040-9383(83)90045-9} {\bibfield
  {journal} {\bibinfo  {journal} {Topology}\ }\textbf {\bibinfo {volume}
  {22}},\ \bibinfo {pages} {47} (\bibinfo {year} {1983})}\BibitemShut {NoStop}%
\bibitem [{\citenamefont {Takens}(2006)}]{takens2006detecting}%
  \BibitemOpen
  \bibfield  {author} {\bibinfo {author} {\bibfnamefont {F.}~\bibnamefont
  {Takens}},\ }\bibfield  {title} {\bibinfo {title} {Detecting strange
  attractors in turbulence},\ }in\ \href@noop {} {\emph {\bibinfo {booktitle}
  {Dynamical Systems and Turbulence, Warwick 1980}}}\ (\bibinfo {organization}
  {Springer},\ \bibinfo {year} {2006})\ pp.\ \bibinfo {pages}
  {366--381}\BibitemShut {NoStop}%
\end{thebibliography}%
\end{document}